\documentclass[twocolumn,showpacs,preprintnumbers,amsmath,amssymb]{revtex4}
\usepackage{graphicx}
\usepackage{bm}
\usepackage{color}
\usepackage{amsmath}
\usepackage{amsfonts}
\usepackage{amssymb}
\usepackage{epstopdf}
\usepackage{ulem}
\usepackage{array}

\begin{document}

\preprint{APS/123-QED}

\title{Quantum magnetic properties of the spin-$\frac{1}{2}$ triangular-lattice antiferromagnet Ba$_2$La$_2$CoTe$_2$O$_{12}$}% Force line breaks with \\

\author{Yuki Kojima$^1$}
\email{kojima.y.ai@m.titech.ac.jp}
\author{Masari Watanabe$^1$}
\author{Nobuyuki Kurita$^1$}
\author{Hidekazu Tanaka$^1$}
\email{tanaka@lee.phys.titech.ac.jp}
\author{Akira Matsuo$^2$}
\author{Koichi Kindo$^2$}
\author{Maxim Avdeev$^{3,4}$}
%email{kojima.y.ai@m.titech.ac.jp}
%email{watanabe.m.bm@m.titech.ac.jp}
%\email{kurita.n.aa@m.titech.ac.jp}
%\email{tanaka@lee.phys.titech.ac.jp}
%\email{a-matsuo@issp.u-tokyo.ac.jp}
%\email{kindo@issp.u-tokyo.ac.jp}
%\email{max@ansto.gov.au}

\affiliation{
$^1$Department of Physics, Tokyo Institute of Technology, Meguro-ku, Tokyo 152-8551, Japan\\
$^2$Institute for Solid State Physics, University of Tokyo, Kashiwa, Chiba 277-8581, Japan\\
$^3$Australian Nuclear Science and Technology Organisation, Lucas Heights, New South Wales 2234, Australia\\
$^4$School of Chemistry, The University of Sydney, Sydney 2006, Australia}
\date{\today}% It is always \today, today,

\begin{abstract}
We report the crystal structure of Ba$_2$La$_2$CoTe$_2$O$_{12}$ determined by Rietveld analysis using x-ray powder diffraction data. It was found from magnetic measurements that Ba$_2$La$_2$CoTe$_2$O$_{12}$ can be described as a spin-$\frac{1}{2}$ triangular-lattice antiferromagnet with easy-plane anisotropy at low temperatures. This compound undergoes a magnetic phase transition at $T_{\rm N}\,{=}\,3.26\,$K to an ordered state with the $120^{\circ}$ structure. The magnetization curve exhibits the one-third plateau characteristic of triangular-lattice quantum antiferromagnets. The antiferromagnetic exchange interaction and the $g$ factors parallel and perpendicular to the $c$ axis were evaluated to be $J/k_{\rm B}\,{=}\,22\,$K, $g_{\parallel}\,{=}\,3.5$ and $g_{\perp}\,{=}\,4.5$, respectively.
\end{abstract}

% PACS, the Physics and Astronomy Classification Scheme.

\maketitle

%--------------------------------------------------------------------------------------------------------------------

\section{Introduction}
A spin-$\frac{1}{2}$ triangular-lattice antiferromagnet (TLAF) is a representative frustrated quantum magnet. Great effort has been made to explore the quantum many-body effect in the spin-$\frac{1}{2}$ TLAFs~\cite{Anderson,Kalmeyer,Balents}. At present, the theoretical consensus is that the ground state of the spin-$\frac{1}{2}$ TLAFs with only the nearest-neighbor isotropic exchange interaction is not a quantum disordered state such as a spin liquid~\cite{Anderson} but an ordered state with the 120$^{\circ}$ spin structure~\cite{Huse,Jolicoeur,Bernu,Singh}. However, the magnitude of the sublattice spins is reduced to approximately 40\% of the full moment owing to the quantum fluctuation~\cite{Capriotti,Zheng,White}. 

Recently, the effect of the exchange randomness in the spin-$\frac{1}{2}$ TLAFs has been investigated~\cite{Watanabe,Shimokawa,Zhu}. It was demonstrated that the spin-liquid-like behavior observed in some molecular magnets \cite{Shimizu,YamashitaS_NatP2008,YamashitaM_NatP2009,Itou1,Itou2} and YbMgGaO$_4$~\cite{Li1,Li2,Shen,Paddison} can be interpreted in terms of the exchange randomness induced by charge disorder~\cite{Nakajima,Watanabe,Shimokawa} and intersite mixing between Mg$^{2+}$ and Ga$^{3+}$~\cite{Li3,Zhu}, respectively.

Although the zero-field ground state of the spin-$\frac{1}{2}$ TLAFs is qualitatively the same as that for the classical spin, several nonclassical spin structures that are unstable in the classical spin model are stabilized in magnetic fields with the help of the quantum fluctuation~\cite{Chubukov,Nikuni,Alicea,Farnell,Honecker,Sakai,Hotta,Yamamoto1,Sellmann,Starykh2,Coletta}. Consequently, the spin-$\frac{1}{2}$ TLAFs undergo magnetic-field-induced quantum phase transitions. The most noticeable quantum effect is that an up-up-down (UUD) spin state, which appears in a magnetic field for the classical model, can be stabilized in a finite magnetic field range, causing a magnetization plateau at one-third of the saturation magnetization~\cite{Chubukov,Nikuni,Alicea,Farnell,Honecker,Sakai,Hotta,Yamamoto1,Sellmann,Starykh2,Coletta}. The 1/3-magnetization plateau of the spin-$\frac{1}{2}$ TLAFs was actually observed in Cs$_2$CuBr$_4$~\cite{Ono1,Ono2,Fortune}, which is a spatially anisotropic spin-$\frac{1}{2}$ TLAF, and quantitatively verified in Ba$_3$CoSb$_2$O$_9$~\cite{Shirata,Susuki,Zhou,Quirion,Koutroulakis}. However, the quantum phases stabilized by the spatial anisotropy~\cite{Ono2,Fortune}, exchange anisotropy~\cite{Yamamoto1} and interlayer exchange interaction~\cite{Susuki,Koutroulakis,Yamamoto2} in magnetic fields have not been sufficiently elucidated.

A remarkable quantum effect has also been predicted for the magnetic excitations. For example, the excitation energy is significantly renormalized downward by quantum fluctuations, causing the dispersion curves to become flat~\cite{Starykh,Zheng,Chernyshev,Mezio,Mourigal,Ghioldi}. The dispersion curve shows a rotonlike minimum at the $M$ point~\cite{Zheng,Ghioldi}. These unusual dynamical properties of the spin-$\frac{1}{2}$ TLAFs were verified by inelastic neutron scattering experiments on Ba$_3$CoSb$_2$O$_9$~\cite{Ma,Ito}. However, the intense dispersive excitation continua extending to a high energy six times the exchange constant cannot be described by the current theory~\cite{Ito}. Thus, the spin-$\frac{1}{2}$ TLAFs include rich quantum many-body effects yet to be fully explained.

Experimentally, a great effort has been made to develop spin-$\frac{1}{2}$ TLAFs. Recently, the magnetic properties of the spin-$\frac{1}{2}$ triangular-lattice magnets Ba$_2$La$_2M$W$_2$O$_{12}$ ($M$\,=\,Mn,\,Co,\,Ni,\,Zn) (where $M$ represents a metal)~\cite{Sack,Rawl,Doi} and Ba$_8$CoNb$_6$O$_{24}$~\cite{Rawl2,Cui} have been reported. A feature of these compounds is that the magnetic triangular lattices are widely separated by the layers of nonmagnetic ions; thus, good two-dimensionality is expected. However, their magnetization data indicate that the exchange interactions are weakly antiferromagnetic~\cite{Rawl,Rawl2,Cui} or weakly ferromagnetic~\cite{Doi}. In these compounds, neighboring spins in the same triangular layer interact via superexchange interactions 
through $M^{2+}$-\,O$^{2-}$-\,O$^{2-}$-\,$M^{2+}$ and $M^{2+}$-\,O$^{2-}$-\,${\rm W}^{6+}({\rm Nb}^{5+})$-\,O$^{2-}$-\,$M^{2+}$ paths. It is considered that the superexchange through the former path is antiferromagnetic, whereas it is ferromagnetic for the latter path, because the filled outermost orbitals of nonmagnetic ${\rm W}^{6+}$ and ${\rm Nb}^{5+}$ ions are $4p$ orbitals, as discussed in Refs.~\cite{Yokota,Koga}. The superexchange interactions via these two paths almost cancel out, resulting in a weakly antiferromagnetic or ferromagnetic total exchange interaction. If nonmagnetic ${\rm W}^{6+}$ and ${\rm Nb}^{5+}$ ions can be replaced by ${\rm Te}^{6+}$ and ${\rm Sb}^{5+}$ ions, respectively, for which the filled outermost orbital is a $4d$ orbital, the superexchange interaction through the $M^{2+}$-O$^{2-}$-${\rm Te}^{6+}({\rm Sb}^{5+})$-\,O$^{2-}$-\,$M^{2+}$ path becomes antiferromagnetic, and the total exchange interaction should be strongly antiferromagnetic. With this motivation, we synthesized Ba$_2$La$_2$CoTe$_2$O$_{12}$. It was found that the structure of this compound is the same as that of Ba$_2$La$_2$CoW$_2$O$_{12}$. Figure~\ref{fig:cryst} shows the crystal structure of Ba$_2$La$_2$CoTe$_2$O$_{12}$. As shown below, the exchange interaction in this compound was found to be $J/k_{\rm B}\,{\simeq}\,22\,$K, which is much larger than those in Ba$_2$La$_2$CoW$_2$O$_{12}$~\cite{Sack,Rawl,Doi} and Ba$_8$CoNb$_6$O$_{24}$~\cite{Rawl2,Cui}. The 1/3-magnetization plateau characteristic of the quantum triangular-lattice antiferromagnets was clearly observed for Ba$_2$La$_2$CoTe$_2$O$_{12}$.

\begin{figure}[t]
\includegraphics[width=6.5cm, clip]{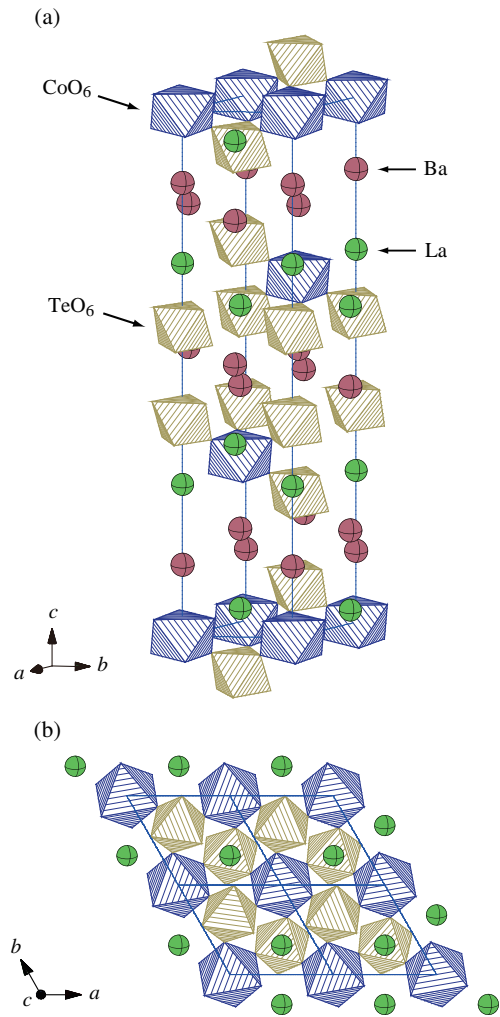}
\caption{(Color online) (a) Schematic of the crystal structure of Ba$_2$La$_2$CoTe$_2$O$_{12}$. The blue and gold single octahedra are CoO$_6$ and TeO$_6$ octahedra centered by Co$^{2+}$ and Te$^{6+}$ ions, respectively.  Solid lines denote the chemical unit cell. (b) Crystal structure viewed along the $c$ axis. Magnetic Co$^{2+}$ ions form a uniform triangular lattice on the $ab$ plane.}
 \label{fig:cryst}
\end{figure} 

Before giving the experimental details, we summarize the effective model of Co$^{2+}$ in an octahedral environment, which is valid below liquid nitrogen temperature. It is known that the magnetic property of Co$^{2+}$ in an octahedral environment is determined by the lowest orbital triplet $^4T_1~$\cite{Abragam,Lines,Shiba}. Within the $^4T_1~$ state, the orbital angular momentum $\bm L$ with $L\,{=}\,3$ can be replaced by $-(3/2){\bm l}$ with $l\,{=}\,1$. This orbital triplet splits into six Kramers doublets owing to the spin-orbit coupling and the uniaxial crystal field, which are expressed together as
\begin{eqnarray}
{\cal H}^{\prime}=-(3/2){\lambda}({\bm l}\cdot{\bm S})-{\delta}\{(l^z)^2-2/3\},
\label{eq:perturb}
\end{eqnarray}
where $\bm S$ is the true spin with $S\,{=}\,3/2$. When the temperature $T$ is much lower than the magnitude of the spin-orbit coupling constant ${\lambda}\,{=}\,{-}178\,$cm$^{-1}$, i.e., $T\,{\ll}\, |{\lambda}|/k_{\rm B}\,{\simeq}\,250$\,K, the magnetic property is determined by the lowest Kramers doublet, which is given by $l^z\,{+}\,S^z\,{=}\,{\pm}1/2$, and the effective magnetic moment of Co$^{2+}$ is represented by ${\bm m}\,{=}\,g{\mu}_{\rm B}{\bm s}$ with the spin-$\frac{1}{2}$ operator $\bm s$~\cite{Abragam,Lines,Shiba}. When the octahedral environment exhibits trigonal symmetry as in Ba$_2$La$_2$CoTe$_2$O$_{12}$, the effective exchange interaction between fictitious spins ${\bm s}_i$ is described by the spin-$\frac{1}{2}$ {\it XXZ} model~\cite{Lines,Shiba}
\begin{eqnarray}
{\cal H}_{\rm ex}=\sum_{<i,j>} \left[J_{\perp}\left\{s_i^xs_j^x+s_i^ys_j^y\right\}+J_{\parallel}s_i^zs_j^z\right].
\label{eq:int}
\end{eqnarray}
This interaction is Ising-like ($J_{\parallel}/J_{\perp}\,{>}\,1$) for ${\delta}/{\lambda}\,{<}\,0$, whereas it is {\it XY}-like ($J_{\parallel}/J_{\perp}\,{<}\,1$) for ${\delta}/{\lambda}\,{>}\,0$. The Heisenberg model ($J_{\parallel}/J_{\perp}\,{=}\,1$) is realized when ${\delta}\,{=}\,0$. In general, the $g$ factor is considerably anisotropic, and the total of the $g$ factors for the three different field directions, $g_{\parallel}+2g_{\perp}$, is about 13~\cite{Abragam}, which is twice as large as that for conventional magnets. As shown below, the CoO$_6$ octahedron in Ba$_2$La$_2$CoTe$_2$O$_{12}$ is compressed along the trigonal axis $c$. In such a case, the exchange interaction becomes $XY$-like and $g_{\perp}\,{>}\,g_{\parallel}$. 

%--------------------------------------------------------------------------------------------------------------------
\section{Experimental Details}
Ba$_2$La$_2$CoTe$_2$O$_{12}$ powder was prepared by the solid-state reaction in accordance with the chemical reaction 2BaCO$_3$\,{+}\,La$_2$O$_3$\,{+}\,CoO\,{+}\,2TeO$_2$\,{+}\,O$_2$ $\longrightarrow$\ Ba$_2$La$_2$CoTe$_2$O$_{12}$\,{+}\,2CO$_2$. BaCO$_3$ (Wako, 99.9\% purity), La$_2$O$_3$ (Wako, 99.99\% purity), CoO (Soekawa, 99.9\% purity) and TeO$_2$ (Sigma-Aldrich, 99.995\% purity) were mixed in stoichiometric quantities and calcined at 1000\,$^{\circ}$C for 24 h in air. Ba$_2$La$_2$CoTe$_2$O$_{12}$ was sintered at 1000\,$^{\circ}$C for 24 h after being pressed into pellets. Dark violet powder samples were obtained.

A powder x-ray diffraction (XRD) measurement was carried out at room temperature using a Rigaku Mini FlexII diffractometer with monochromatized Cu\,$K\alpha$ radiation. Rietveld refinement of the powder XRD data was carried out by the RIETAN-FP program \cite{Izumi2007}. The obtained powdered sample was confirmed by XRD measurement to have the same crystal structure as Ba$_2$La$_2M$W$_2$O$_{12}$ ($M$\,=\,Mn,\,Co,\,Ni,\,Zn)~\cite{Sack,Rawl,Doi}, as shown below.

\begin{figure}[t]
	\centering
	\includegraphics[width=8.5cm, clip]{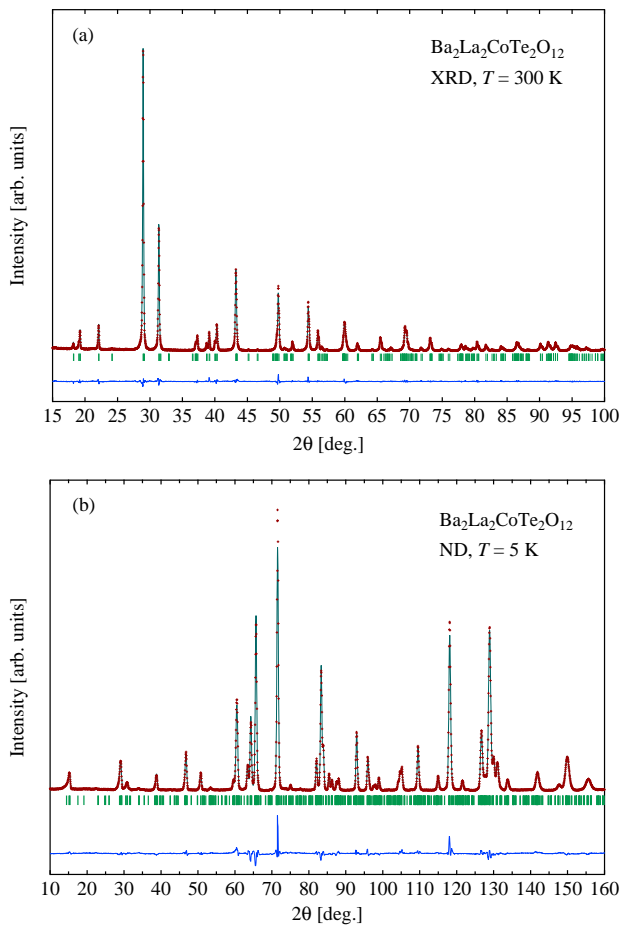}
	\caption{(Color online) (a) XRD pattern and (b) ND pattern of Ba$_2$La$_2$CoTe$_2$O$_{12}$ measured at room temperature and $T\,{=}\,5\,$K, respectively. Experimental data, the results of Rietveld fitting and their difference are shown by the red symbols, green line and blue curves, respectively.}
	\label{fig:xrd}
\end{figure}

The magnetic susceptibility of Ba$_2$La$_2$CoTe$_2$O$_{12}$ powder was measured in the temperature range of $1.8$-$300\,$K using a superconducting quantum interference device (SQUID) magnetometer (MPMS XL, Quantum Design).
High-field magnetization was measured  in a magnetic field of up to 60\,T at 1.3\,K using an induction method with a multilayer pulse magnet at  the Institute for Solid State Physics (ISSP), The University of Tokyo. The absolute value of the high-field magnetization was calibrated with the magnetization measured using the SQUID magnetometer.
The specific heat of Ba$_2$La$_2$CoTe$_2$O$_{12}$ powder was measured down to 0.4\,K in a magnetic field of up to 9\,T using a physical property measurement system (PPMS, Quantum Design) by the relaxation method.

Powder neutron diffraction (ND) measurement of Ba$_2$La$_2$CoTe$_2$O$_{12}$ was performed using the high-resolution powder diffractometer Echidna installed at the OPAL reactor of the Australian Nuclear Science and Technology Organisation (ANSTO). The diffraction data were collected with a neutron wavelength of 2.4395 $\rm{\AA}$. Rietveld refinement of the powder ND data was performed by the FULLPROF program~\cite{Carvajal}.

\section{Results and Discussion}
\subsection{Crystal structure}
%angle
\begin{table}[t]
\begin{center}
  \renewcommand{~}{\phantom{$-$}}
  \caption{Structural parameters of Ba$_2$La$_2$CoTe$_2$O$_{12}$ at room temperature determined by powder XRD measurement.}
  \label{XRD}
  {\setlength{\tabcolsep}{8.5pt}
    \begin{tabular}{lllll}
      \hline
\multicolumn{1}{c}{Atom}& \multicolumn{1}{c}{$x$} & \multicolumn{1}{c}{$y$} &  \multicolumn{1}{c}{$z$}&  $B$ $(\rm{\AA}^2)$\\
      \hline
      Ba &  0 & 0 & 0.13680(3) & 0.34(2)\\
      La &  0 & 0 & 0.28941(4) & 0.24(3)\\
      Co &  0 & 0 & 0 & 0.18(8)\\
      Te &  0 & 0 & 0.41531(4) & 0.20(2)\\
      O1 &  0.459(2) & 0.457(2) & 0.1177(2) & 0.6(1)\\
      O2 &  0.441(2) & 0.462(2) & 0.2954(2) & $B$(O1)\\
      \hline
      \multicolumn{5}{l}{\vspace{-3mm}\,}\\
      \multicolumn{5}{l}{\vspace{0.3mm}Space group $R{\bar 3}$}\\
      \multicolumn{5}{l}{\vspace{0.3mm}$a=5.693(2)\,\rm{\AA},\ c=27.585(6)\,\rm{\AA}$}\\
      \multicolumn{5}{l}{\vspace{0.3mm}$R_{\rm wp}=8.17\%,\ R_{\rm p}=5.87\%,\ R_{\rm e}=7.15\%,$}\\
      \multicolumn{5}{l}{\vspace{0.3mm}$R_{\rm B}=2.18\%,\ R_{\rm F}=1.21\%$}\\
      \hline
    \end{tabular}
  }
  \end{center}
\end{table}
\begin{table}
\begin{center}
  \renewcommand{~}{\phantom{$-$}}
  \caption{Structural parameters of Ba$_2$La$_2$CoTe$_2$O$_{12}$ at $T\,{=}\,5$\,K determined by the powder ND measurement.}
  \label{ND}
  {\setlength{\tabcolsep}{13pt}
 \begin{tabular}{llll}
      \hline
      \multicolumn{1}{c}{Atom} & \multicolumn{1}{c}{$x$} & \multicolumn{1}{c}{$y$} &  \multicolumn{1}{c}{$z$}\\
      \hline
      Ba & 0 & 0 & 0.1366(1) \\
      La & 0 & 0 & 0.28920(6) \\
      Co & 0 & 0 & 0 \\
      Te & 0 & 0 & 0.4149(1) \\
      O1 & 0.4605(3) & 0.4646(4) & 0.11614(4) \\
      O2 & 0.4298(3) & 0.4608(4) & 0.29509(5) \\
      \hline
      \multicolumn{4}{l}{\vspace{-3mm}\,}\\
      \multicolumn{4}{l}{\vspace{0.3mm}Space group $R{\bar 3}$}\\
      \multicolumn{4}{l}{\vspace{0.3mm}$a=5.693(1)\,\rm{\AA},\ c=27.517(4)\,\rm{\AA}$}\\
      \multicolumn{4}{l}{\vspace{0.3mm}Overall $B$ factor $=0.545(3)\,\rm{\AA}^2$}\\
      \multicolumn{4}{l}{\vspace{0.3mm}$R_{\rm wp}=10.6\%,\ R_{\rm p}=8.37\%,\ R_{\rm e}=4.29\%$}\\
      \hline
    \end{tabular}
  }
  \end{center}
\end{table}

The XRD pattern of Ba$_2$La$_2$CoTe$_2$O$_{12}$ measured at room temperature and the result of Rietveld analysis are shown in Fig.~\ref{fig:xrd}(a). The analysis was based on two structural models with space groups $R\bar{3}m$ and $R\bar{3}$. Because both structural models successfully reproduce the observed XRD pattern, it is difficult to determine the space group from only the XRD pattern. However, the neutron diffraction pattern obtained at $T\,{=}\,5\,$K is much better described by space group $R\bar{3}$, i.e., the $R_{\rm wp}$-factors for $R\bar{3}$ and $R\bar{3}m$ are $R_{\rm wp}\,{=}\,10.6$ and 38.3\%, respectively. Figure~\ref{fig:xrd}(b) shows the ND pattern measured at 5\,K and the result of Rietveld analysis. From the magnetic susceptibility and specific heat data, no structural phase transition was observed down to liquid helium temperatures. Thus, we refined the crystal structure using the structural model with space group $R\bar{3}$, as in the cases of Ba$_2$La$_2M$W$_2$O$_{12}$ ($M$\,=\,Mn, Co, Ni, Zn)~\cite{Doi}. The refined structural parameters at room temperature and $T\,{=}\,5\,$K are listed in Tables \ref{XRD} and \ref{ND}, respectively. With decreasing temperature, the lattice constant $c$ is decreases, whereas the lattice constant $a$ is almost independent of temperature.

The refined crystal structure of Ba$_2$La$_2$CoTe$_2$O$_{12}$ is shown in Fig.~\ref{fig:cryst}. CoO$_6$ and TeO$_6$ octahedra rotate around the $c$ axis; thus, mirror symmetry parallel to the $c$ axis is absent.
Magnetic Co$^{2+}$ ions form a uniform triangular lattice parallel to the $ab$ plane. The lattice point of one triangular lattice shifts onto the center of the triangle of the neighboring triangular lattices when viewed along the $c$ axis. Consequently, the triangular layers are stacked as $ABCABC\cdots$ along the $c$ axis. Because of the inversion symmetry at the center of neighboring Co$^{2+}$ ions, antisymmetric interaction of the Dzyaloshinskii--Moriya type is absent between neighboring spins. Since the magnetic triangular layers are widely separated by nonmagnetic layers, it is expected that the interlayer exchange interaction is much smaller than the intralayer exchange interaction. The CoO$_6$ octahedra are trigonally compressed, which leads to the condition $J_{\perp}\,{>}\,J_{\parallel}$ in Eq.~(\ref{eq:int})~\cite{Lines}. Thus, we deduce that Ba$_2$La$_2$CoTe$_2$O$_{12}$ has the exchange anisotropy of easy-plane type, as in the case of Ba$_3$CoSb$_2$O$_9$~\cite{Susuki,Koutroulakis,Quirion,Ma}.

\subsection{Magnetic susceptibility}
\begin{figure}[t]
	\centering
	\includegraphics[width=8.5cm, clip]{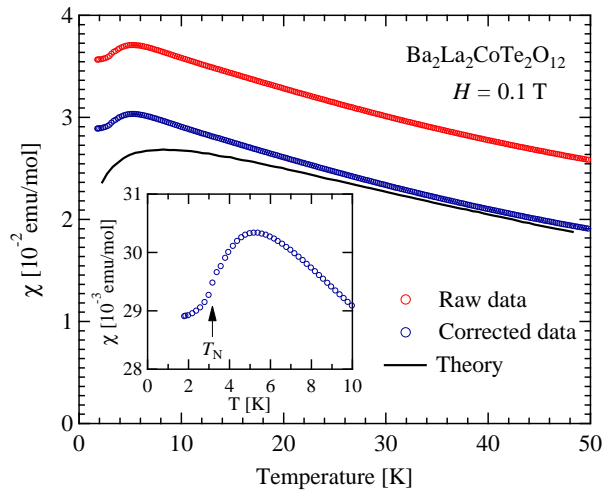}
	\caption{(Color online) Temperature dependence of the magnetic susceptibility of Ba$_2$La$_2$CoTe$_2$O$_{12}$ obtained before and after correction of Van Vleck paramagnetism. The solid line denotes the theoretical susceptibility calculated by series expansion~\cite{Elstner} with $J/k_{\rm B}=22\,$K and $g=4.2$.
	}
	\label{fig:MT}
\end{figure}

Figure~\ref{fig:MT} shows the temperature dependence of the magnetic susceptibility of Ba$_2$La$_2$CoTe$_2$O$_{12}$ powder obtained before and after the correction of the Van Vleck paramagnetic susceptibility of $\chi_{\rm VV}\,{=}\,6.75\,{\times}\,10^{-3}\,$emu/mol, which was evaluated from the magnetization slope above the saturation field shown in Fig.~\ref{fig:MH}(a). We plotted the susceptibility data for $T\,{\leq}\,50\,$K, where the spin-$\frac{1}{2}$ description of the magnetic moment is valid. The inset of Fig.~\ref{fig:MT} shows an enlargement of the susceptibility data below 10\,K. With decreasing temperature, the magnetic susceptibility exhibits a rounded maximum at approximately 5\,K, which is characteristic of a low-dimensional antiferromagnet. With further decreasing temperature, the susceptibility has an inflection point at $T_{\rm N}\,{=}\,3.2$\,K, which can be assigned to the transition temperature of three-dimensional magnetic ordering. 

The solid line in Fig.~\ref{fig:MT} indicates the theoretical susceptibility of the spin-$\frac{1}{2}$ Heisenberg TLAF calculated by series expansion~\cite{Elstner} with $J/k_{\rm B}\,{=}\,22\,$K and $g\,{=}\,4.2$, which were obtained from the present high-field magnetization measurements, as shown below. Above 30\,K, the experimental and theoretical susceptibilities coincide. However,  below 30\,K, the experimental susceptibility deviates from the theoretical result. This deviation is considered to be caused by the {\it XY}-like exchange anisotropy in Ba$_2$La$_2$CoTe$_2$O$_{12}$. The exchange interaction in Ba$_2$La$_2$CoTe$_2$O$_{12}$ is antiferromagnetic and its magnitude is much larger than those in Ba$_2$La$_2M$W$_2$O$_{12}$ ($M$\,=\,Mn, Co, Ni, Zn)~\cite{Rawl,Rawl2,Cui,Doi}. This result demonstrates that the filled outermost orbitals of nonmagnetic hexavalent ions play a crucial role in the superexchange interaction.

\subsection{Magnetization process}
\begin{figure}[t]
	\centering
	\includegraphics[width=8.5cm, clip]{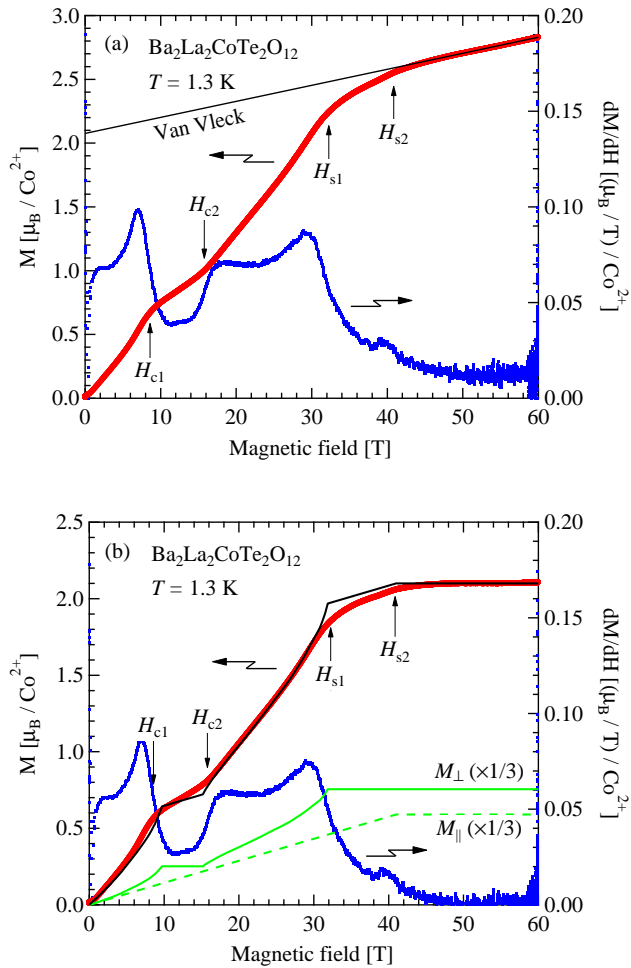}
	\caption{(Color online) (a) Raw magnetization curve of Ba$_2$La$_2$CoTe$_2$O$_{12}$ powder and derivative susceptibility $dM/dH$ vs magnetic field $H$ measured at 1.3\,K. The critical fields $H_{\rm c1}$ and $H_{\rm c2}$ are the lower- and upper-edge fields, respectively, of the 1/3-magnetization plateau for $H\,{\perp}\,c$. $H_{\rm s1}$ and $H_{\rm s2}$ are the saturation fields for $H\,{\perp}\,c$ and $H\,{\parallel}\,c$, respectively. The solid line denotes the Van Vleck paramagnetism evaluated from the magnetization slope above the saturation field $H_{\rm s2}\,{=}\,41\,$T. (b) Magnetization curve corrected for Van Vleck paramagnetism. The green solid line is the theoretical magnetization curve $M_{\perp}$ for $H\,{\perp}\,c$ calculated by the higher-order coupled cluster method (CCM) for an $S\,{=}\,1/2$ Heisenberg TLAF~\cite{Farnell} and the green dashed line is the linear magnetization curve $M_{\parallel}$ assumed for $H\,{\parallel}\,c$. The black solid line denotes the total magnetization given by $M_{\rm tot}\,{=}\,(2M_{\perp}+M_{\parallel})/3$.}
	\label{fig:MH}
\end{figure}

Figure~\ref{fig:MH}(a) shows the raw magnetization curve and derivative susceptibility $dM/dH$ vs the magnetic field $H$ of the Ba$_2$La$_2$CoTe$_2$O$_{12}$ powder measured at 1.3\,K. The saturation of the Co$^{2+}$ spin starts at $H_{\rm s1}\,{=}\,32\,$T and ends at $H_{\rm s2}\,{=}\,41\,$T. The distribution of the saturation field is attributed to the anisotropy of the $g$ factor and the exchange anisotropy of the easy-plane type, which are produced by the trigonal compression of the CoO$_6$ octahedron, as discussed in Section I. The linear increase in magnetization above $H_{\rm s2}$ is ascribed to the large Van Vleck paramagnetism characteristic of Co$^{2+}$ in the octahedral environment. The Van Vleck paramagnetic susceptibility is evaluated as $\chi_{\rm VV}\,{=}\,1.21\times10^{-2}\,{\rm(\mu_{\rm B}/T)/Co^{2+}}\,{=}\,6.75\times 10^{-3}\,$emu/mol, which is somewhat smaller than $\chi_{\rm VV}\,{=}\,8.96\,$emu/mol observed for Ba$_3$CoSb$_2$O$_9$~\cite{Shirata}. The saturation magnetization was obtained as $M_{\rm s}\,{=}\,2.10\,\mu_{\rm B}/$Co$^{2+}$ by extrapolating the magnetization curve above $H_{\rm s2}$ to a zero field, as shown by the solid line in Fig.~\ref{fig:MH}(a). From the saturation magnetization, the average $g$ factor is evaluated to be $g_{\rm avg}\,{=}\,4.2$.

Figure~\ref{fig:MH}(b) shows the magnetization curve corrected for the contribution of Van Vleck paramagnetism. The magnetization curve exhibits a magnetization plateau at one-third of the saturation magnetization, which is characteristic of quantum TLAFs. When the CoO$_6$ octahedron is compressed trigonally, as observed in Ba$_2$La$_2$CoTe$_2$O$_{12}$, the $g$ factor $g_{\perp}$ for $H\,{\perp}\,c$ is greater than $g_{\parallel}$ for $H\,{\parallel}\,c$~\cite{Abragam,Lines}. Thus, we can deduce that $H_{\rm s1}$ and $H_{\rm s2}$ are the saturation fields for $H\,{\perp}\,c$ and $H\,{\parallel}\,c$, respectively. Neglecting the anisotropy of the saturation field due to the exchange anisotropy, the relation between the saturation fields and the $g$ factors is given by $g_{\perp}{\mu}_{\rm B}H_{\rm s1}\,{=}\,g_{\parallel}{\mu}_{\rm B}H_{\rm s2}\,{=}\,4.5J$. Using this equation and the relation $g_{\rm avg}=(2g_{\perp}+g_{\parallel})/3$ together with $H_{\rm s1}\,{=}\,32,\,H_{\rm s2}\,{=}\,41\,$T and $g_{\rm avg}\,{=}\,4.2$, we obtain $g_{\perp}\,{=}\,4.5$, $g_{\parallel}\,{=}\,3.5$ and $J/k_{\rm B}\,{=}\,22\,$K. This exchange constant is close to $J/k_{\rm B}\,{=}\,20.5\,$K for subsystem $A$ in Ba$_2$CoTeO$_6$~\cite{Nick}, where the paths of the superexchange are similar to those in Ba$_2$La$_2$CoTe$_2$O$_{12}$.

\begin{figure}[t]
	\centering
	\includegraphics[width=7.5cm, clip]{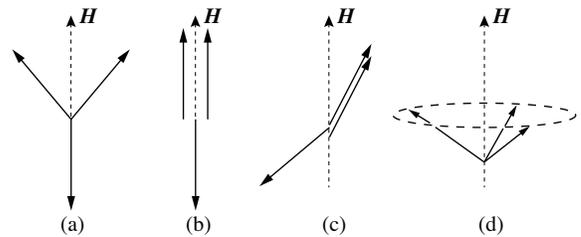}
	\caption{(Color online) Spin structures of two-dimensional triangular-lattice antiferromagnet in the magnetic field. (a) Triangular coplanar structure, (b) UUD structure, (c) high-field coplanar structure, and (d) umbrella structure.}
	\label{fig:spin_structure}
\end{figure}

%theoritical
The 1/3-magnetization plateau shown in Fig.~\ref{fig:MH}(b) is not flat but has a finite slope. In the case where the exchange anisotropy is of the easy-plane type and the interlayer exchange interaction is antiferromagnetic, the magnetization curve exhibits the 1/3-magnetization plateau when the field direction is roughly parallel to the triangular layer, whereas when the field direction is nearly perpendicular to the triangular layer, the magnetization curve exhibits no plateau as observed for Cs$_2$CuBr$_4$~\cite{Ono1,Ono2} and Ba$_3$CoSb$_2$O$_9$~\cite{Susuki,Koutroulakis,Yamamoto2}. For this reason, the 1/3-magnetization plateau has a finite slope for powder-averaged data. 

To analyze the magnetization curve, we compare our experimental result with a simplified magnetization curve that is given by a linear combination of the model magnetization curves for $H\,{\perp}\,c$ and $H\,{\parallel}\,c$. As the model magnetization curve $M_{\perp}$ for $H\,{\perp}\,c$, we use the result calculated by the higher-order coupled cluster method (CCM) for an $S\,{=}\,1/2$ Heisenberg TLAF~\cite{Farnell} as shown by the green solid line in Fig.~\ref{fig:MH}(b). In this field direction, the triangular coplanar, UUD and high-field coplanar structures, respectively illustrated in Figs.~\ref{fig:spin_structure}(a)-\ref{fig:spin_structure}(c), emerge with increasing magnetic field. For $H\,{\parallel}\,c$, we assume that the magnetization $M_{\parallel}$ increases linearly with the magnetic field up to the saturation as shown by the green dashed line in Fig.~\ref{fig:MH}(b). In this field direction, the transition from the umbrella structure [Fig.~\ref{fig:spin_structure}(d)] to the high-field coplanar structure occurs with a small magnetization jump~\cite{Susuki,Koutroulakis}. However, we neglect this small magnetization anomaly for simplification. The black solid line in Fig.~\ref{fig:MH}(b) is the total magnetization given by $M_{\rm tot}=(2M_{\perp}+M_{\parallel})/3$. The field dependence of $M_{\rm tot}$ captures the features of the experimental magnetization curve. The experimental field range of the 1/3-magnetization plateau $H_{\rm c2}\,{-}\,H_{\rm c1}\,{=}\,7.2\,$T is somewhat larger than the calculated field range of 5.5\,T. We infer that this result is caused by the {\it XY}-like anisotropy of the exchange interaction, which decreases the lower-edge field $H_{\rm c1}$~\cite{Chubukov}. 

\begin{figure}[t]
	\centering
	\includegraphics[width=8.5cm, clip]{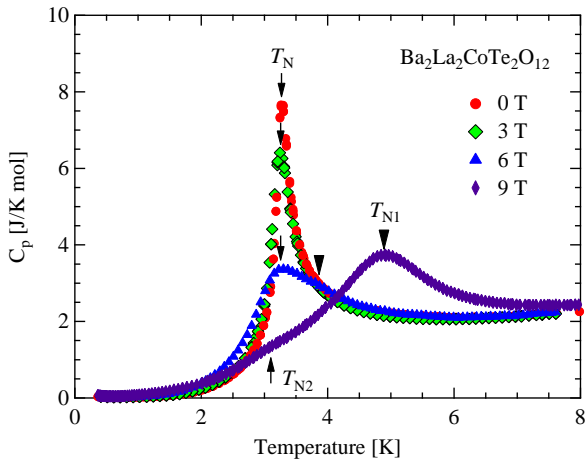}
	\caption{(Color online) Temperature dependence of specific heat of Ba$_2$La$_2$CoTe$_2$O$_{12}$ powder measured at several magnetic fields below 8\,K. Arrows and inverted triangles denote magnetic phase transition temperatures.}
	\label{fig:capacity}
\end{figure}  

\subsection{Low-temperature specific heat}

Figure \ref{fig:capacity} shows the temperature dependence of the specific heat of the Ba$_2$La$_2$CoTe$_2$O$_{12}$ powder measured in magnetic fields of $H\,{=}\,0$, 3, 6, and 9\,T. At zero magnetic field, the specific heat shows a single sharp peak at $T_{\rm N}\,{=}\,3.26\,$K indicative of three-dimensional magnetic ordering. This transition temperature is slightly lower than $T_{\rm N}\,{=}\,3.8\,$K for Ba$_3$CoSb$_2$O$_9$~\cite{Shirata,Doi2}, although the exchange interaction $J/k_{\rm B}\,{=}\,22\,$K for Ba$_2$La$_2$CoTe$_2$O$_{12}$ is somewhat larger than $J/k_{\rm B}\,{=}\,18.2\,$K for Ba$_3$CoSb$_2$O$_9$. This result indicates good two-dimensionality in Ba$_2$La$_2$CoTe$_2$O$_{12}$. The ordered phase below $T_{\rm N}\,{=}\,3.26\,$K has the 120$^{\circ}$ spin structure as shown below. 

With increasing magnetic field, the peak height decreases. At $H\,{=}\,6\,$T, the specific heat has a rounded peak at $T_{\rm N2}(6\,{\rm T})\,{=}\,3.2\,$K and a weak shoulder anomaly at $T_{\rm N1}(6\,{\rm T})\,{=}\,3.8\,$K. The smearing of the transition arises from the random distribution of the principal axes in the powdered sample. At $H\,{=}\,9\,$T, the specific heat has a weak shoulder anomaly at $T_{\rm N2}(9\,{\rm T})\,{=}\,3.1\,$K and a rounded peak at $T_{\rm N1}(9\,{\rm T})\,{=}\,4.9\,$K. By referring to the phase diagram of Ba$_3$CoSb$_2$O$_9$~\cite{Quirion}, which has weak easy-plane anisotropy, we infer that the phase transitions at $T_{\rm N1}$ and $T_{\rm N2}$ above 6\,T correspond to the transitions from the paramagnetic phase to the UUD phase for $H\,{\perp}\,c$ and from the UUD phase to the triangular coplanar phase, respectively.

\subsection{Spin structure in the ordered phase}
\begin{figure}[t]
	\centering
	\includegraphics[width=8.5cm, clip]{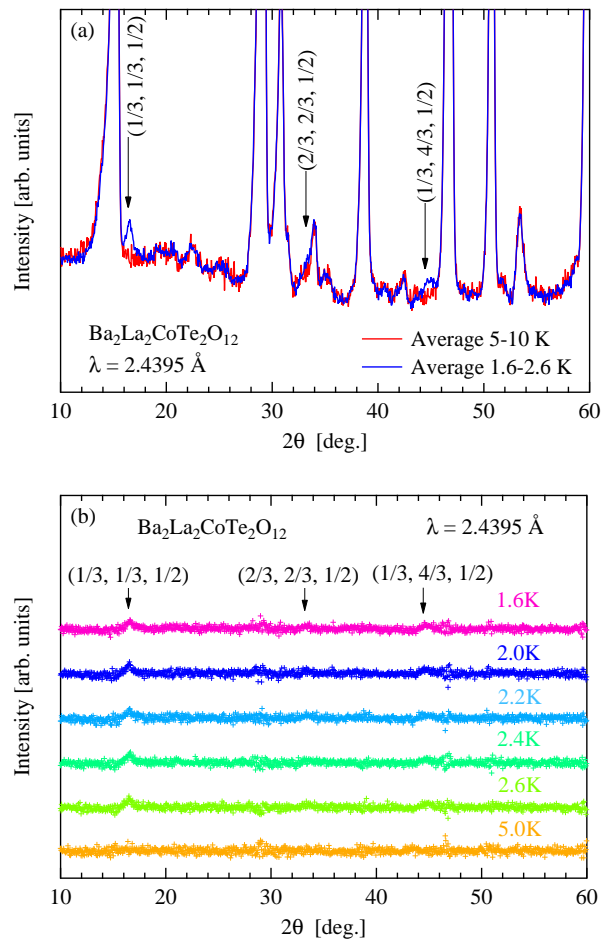}
	\caption{(Color online) (a) Neutron powder diffraction patterns of Ba$_2$La$_2$CoTe$_2$O$_{12}$ averaged over 5-10\,K ($\,{>}\,T_{\rm N}\,{=}\,3.26\,$K) and over 1.6-2.6\,K ($\,{<}\,T_{\rm N}$). Three weak magnetic Bragg peaks assigned to ${\bm Q}_{\rm m}\,{=}\,(1/3,1/3,1/2)$, $(2/3,2/3,1/2)$ and $(1/3,4/3,1/2)$ are observed below $T_{\rm N}$. (b) Neutron powder diffraction spectra collected at various temperatures, where the average of the diffraction spectra collected over 5-10\,K was subtracted as the background. }
	\label{fig:ND}
\end{figure}

To confirm the spin structure in the ordered phase of Ba$_2$La$_2$CoTe$_2$O$_{12}$, we carried out powder ND measurements. Figure~\ref{fig:ND}(a) shows diffraction patterns averaged over 5-10\,K ($\,{>}\,T_{\rm N}\,{=}\,3.26$\,K) and over 1.6-2.6\,K ($\,{<}\,T_{\rm N}$), which are displayed by red and blue lines, respectively. Figure~\ref{fig:ND}(b) shows powder ND spectra obtained at various temperatures, where the average of the diffraction spectra obtained over $5\,{-}\,10$\,K was subtracted as the background. Three weak magnetic Bragg peaks are observed below $T_{\rm N}\,{=}\,3.26\,$K. These magnetic peaks can be assigned to $\bm Q_{\rm m}\,{=}\,(1/3, 1/3, 1/2)$, $(2/3, 2/3, 1/2)$ and $(1/3, 4/3, 1/2)$ on the unit cell of the hexagonal representation. This result indicates that the magnetic structure below $T_{\rm N}$ is the 120$^{\circ}$ structure characterized by the propagation vector ${\bm k}\,{=}\,(1/3, 1/3, 1/2)$. 

From the Rietveld analysis, it was found that spins lie on the $ab$ plane and form the 120$^{\circ}$ structure. Because the lattice point of one triangular lattice shifts onto the center of the triangle of the neighboring triangular lattices when viewed along the $c$ axis, the molecular fields from the neighboring layers cancel out. We investigated the effect of quantum fluctuation on the spin structure on the basis of linear spin wave theory. We found that the quantum fluctuation stabilizes a spin structure with the propagation vector ${\bm k}\,{=}\,(1/3, 1/3, 1/2)$ when the interlayer exchange interaction is ferromagnetic, while for the antiferromagnetic interlayer interaction, a spin structure with the propagation vector ${\bm k}\,{=}\,(1/3, 1/3, 0)$ is preferred. Thus, we deduce that the interlayer exchange interaction is ferromagnetic in Ba$_2$La$_2$CoTe$_2$O$_{12}$. 

The magnitude of the ordered moment was found to be $m\,{=}\,0.63{\mu}_{\rm B}$ at $T\,{=}\,1.6$\,K. Using $g_{\perp}\,{=}\,4.5$, the size of the sublattice spin was estimated to be $\langle S\rangle\,{=}\,0.14$, which is somewhat smaller than $\langle S\rangle\,{=}\,0.20$ calculated for a spin-$\frac{1}{2}$ Heisenberg TLAF~\cite{Capriotti,Zheng,White}. The difference is considered to be due to the frustration of interlayer exchange interactions and the finite temperature effect.

\section{Conclusion}
We have presented the results of a structural analysis and magnetization, specific-heat, and neutron-diffraction measurements on Ba$_2$La$_2$CoTe$_2$O$_{12}$ powder. From the neutron powder-diffraction data, the crystal structure was determined to be trigonal $R{\bar 3}$, which is the same as those of Ba$_2$La$_2M$W$_2$O$_{12}$ ($M$\,=\,Mn,\,Co,\,Ni,\,Zn)~\cite{Doi}. However, different from the weak antiferromagnetic or ferromagnetic exchange interactions observed in these tungsten compounds, the exchange interaction in Ba$_2$La$_2$CoTe$_2$O$_{12}$ was found to be antiferromagnetic and large, $J/k_{\rm B}\,{=}\,22$\,K, as expected from the superexchange path via the filled outermost orbital of nonmagnetic ${\rm Te}^{6+}$~\cite{Yokota,Koga}. Ba$_2$La$_2$CoTe$_2$O$_{12}$ undergoes three-dimensional magnetic ordering at $T_{\rm N}\,{=}\,3.26\,$K. The spin structure  in the ordered phase is characterized by the propagation vector ${\bm k}\,{=}\,(1/3, 1/3, 1/2)$. The spins form the $120^{\circ}$ structure parallel to the triangular layer. The magnetization curve displays a clear 1/3-magnetization plateau, which is characteristic of quantum TLAFs. From the analysis of the magnetization curve, we deduce that the 1/3-plateau emerges for $H\,{\perp}\,c$ but not for $H\,{\parallel}\,c$ owing to the reasonably large exchange anisotropy of the easy-plane type. 
It was also found that the saturation field is anisotropic owing to the anisotropy of the $g$ factor. From the saturation field and saturation magnetization, the $g$ factors and exchange constant were estimated as $g_{\perp}\,{=}\,4.5$, $g_{\parallel}\,{=}\,3.5$ and $J/k_{\rm B}\,{=}\,22\,$K, respectively. Ba$_2$La$_2$CoTe$_2$O$_{12}$ is thus magnetically described as an $S\,{=}\,1/2$ quasi-two-dimensional TLAF with XY-like anisotropy.\\

\section*{Acknowledgments}
This work was supported by Grants-in-Aid for Scientific Research (A) (No.~17H01142) and (C) (No.~16K05414) from Japan Society for the Promotion of Science.

\end{document}